\documentclass[]{raa}            
\usepackage{graphicx,times}
\usepackage{amssymb}
\usepackage{natbib}

\begin{document}
   \title{The optical/UV excess of isolated neutron stars in the RCS model
}

 \volnopage{ {\bf 2009} Vol.\ {\bf 9} No. {\bf XX}, 000--000}
   \setcounter{page}{1}

   \author{H. Tong
      \inst{1}
   \and R. X. Xu
      \inst{2}
      \and L. M. Song
      \inst{1}}

   \institute{ Institute of High Energy Physics, Chinese Academy of Sciences,
                Beijing 100049, China; {\it haotong@ihep.ac.cn}\\
        \and
             School of Physics and State Key
Laboratory of Nuclear Physics and Technology, Peking University, Beijing 100871, China
                     \\\vs \no
   {\small Received [year] [month] [day]; accepted [year] [month] [day] }
}

\abstract{The X-ray dim isolated neutron stars (XDINSs) are peculiar pulsar-like objects,
characterized by their very well Planck-like spectrum. In studying their spectral energy distributions,
the optical/UV excess is a long standing problem. Recently, Kaplan et al. (2011) have measured the optical/UV excess
for all seven sources, which is understandable in the resonant cyclotron scattering (RCS) model previously addressed.
The RCS model calculations show that the RCS process can account for the observed optical/UV excess for most sources .
The flat spectrum of RX J2143.0+0654 may due to contribution from bremsstrahlung emission of the electron system in addition to the RCS process.
 \keywords{pulsars: individual: (RX J0420.0-5022, RX J0720.4-3125, RX J0806.4-4123,
RX J1308.6+2127, RX J1605.3+3249, RX J1856.5-3754,
RX J2143.0+0654), stars: neutron} }

   \authorrunning{Tong, Xu \& Song}            
   \titlerunning{Origin of optical/UV excess in XDINSs}  
   \maketitle


%
%

\section{Introduction}

The seven X-ray dim isolated neutron stars (XDINSs) are puzzling
pulsar-like compact stars. They are characterized by their very well Planck-like spectrum, and then
they are good specimen for neutron star cooling, atmospheric and equation of state studies
(Tong et al. 2007,2008; review Turolla 2009). Additionally, these sources may also have some relations with other classes of pulsar-like objects, e.g.,
magnetars and rotating radio transients, etc (Tong et al. 2010).

The spectral energy distributions of XDINSs show that their optical and UV emissions are above the extrapolation
of the X-ray blackbodies, i.e., optical/UV excess. This excess of XDINSs is a long standing problem which may be related
to the equation of state of neutron stars (Xu 2002, 2009). Previously, only two sources (RX J1856.5-3754 and RX J0720.4-3125)
have both optical and UV data (van Kerkwijk \& Kaplan 2007). Recently, Kaplan et al. (2011) have presented results of the optical and UV
emission for all seven sources. They also find that the source RX J2143.0+0654 has a very flat spectrm
($F_{\nu} \propto \nu^{\alpha}$, $\alpha \sim 0.5$). This new observation requires updated modeling of existing theoretical models.

Tong et al. (2010) have considered the resonant cyclotron scattering (RCS) in pulsar magnetospheres as the origin of optical/UV excess in
XDINSs, i.e., the RCS model. This model can explain the observations of RX J1856.5-3754 and RX J0720.4-3125 (Tong et al. 2010).
Considering the recent observational progress, we apply the RCS model to all seven XDINSs in this paper.

Section 2 is summary of the RCS model. Its application to XDINSs is treated in Section 3. 
Discussions and conclusions are presented in section 4.

\section{The RCS model: a summary}

Tong et al. (2010) have considered the RCS model for the origin of optical/UV excess in XDINSs. It has the following three key points.
\begin{enumerate}
 \item In analogy with magnetospheric study of magnetars and rotating radio transients, also similar to the ``electron blanket''
of neutron stars (Ruderman 2003), there may also be a dense electron system in closed field line regions of XDINSs.
We call it the ``pulsar inner radiation belt'' (since it is not far away from the central star), in comparison with the ``pulsar radiation belt''
proposed by Luo \& Melrose (2007)
(It may be called the ``pulsar outer radiation belt'', since it is near the light cylinder far away from the central star).

\item The scatterings between surface X-ray photons and the electron system are modeled three dimensionally using the Kompaneets equation
method. Numerical calculations show that there exist both down scattering and up scattering of the RCS process. The existence of down scattering
of RCS process is crucial for its application to XDINSs.

 \item  The application of down scattering of RCS process to XDINSs can explain its optical/UV excess problem.
\end{enumerate}

The Kompaneets equation (photon diffusion equation) for RCS process is (Tong et al. 2010):
\begin{equation}\label{The Kompaneets eq. using optical depth}
\left ( \frac{\partial n}{\partial t}\right )_{\mathrm{RCS}} =
\frac{k T_e}{m_e c^2} \frac{1}{x^2} \frac{\partial}{\partial x}
\left\{x^4 \frac{\tau_{\mathrm{RCS}}}{r_1-r_2} c\, g_{\theta} \left[
\frac{\partial n}{\partial x} + n (1+n) \right] \right\}.
\end{equation}
Here, $n$ is the photon occupation number, its evolution with time
(which corresponds to spectra change) is determined by the right
hand side of equation (\ref{The Kompaneets eq. using optical
depth}). $x$ is the dimensionless photon frequency in units of
electron temperature, $\tau_{\mathrm{RCS}}$ is the RCS optical
depth. In order to calculate the RCS modified blackbody spectrum,
four input parameters are needed: the neutron star surface
temperature, the temperature and density of the electron system, and
a normalization constant.

For photons in a certain frequency range, e.g., from optical to soft X-ray,
only electrons in a specific radius range will contribute to the RCS process. The electron system may extend to much larger radius,
however the electrons there will not contribute to the RCS process (Tong et al. 2010). On the other hand, the electrons will radiate thermal
bremsstrahlung emission, which indeed depends on the radial extension of the electron system. The inclusion of thermal bremsstrahlung process
in addition to the RCS process will result in a flat optical spectrum.

\section{Application of the RCS model to X-ray dim isolated neutron stars}

\begin{figure}[t]
\centering
  \includegraphics[width=0.6\textwidth]{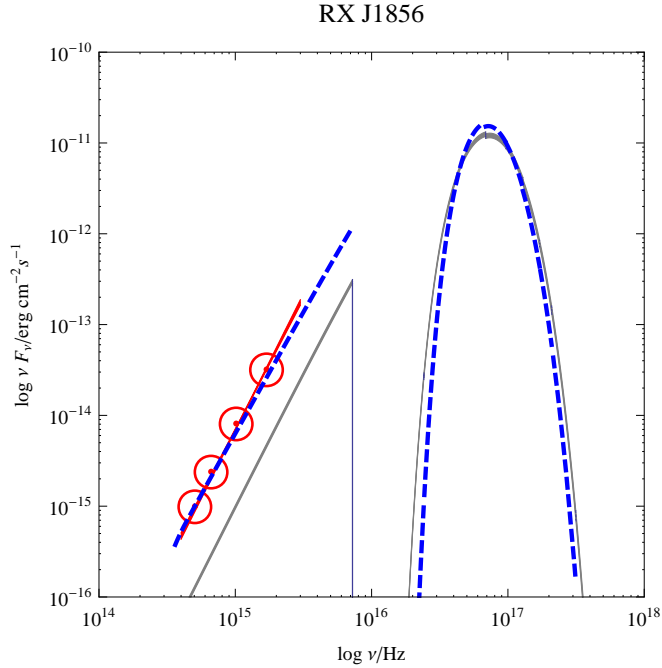}
  \caption{Spectral energy distributions of RX J1856.5-3754 and RCS model calculations.
The gray region is the X-ray blackbody and its extrapolation to the
optical and UV range, including 1-$\sigma$ uncertainties. The red
region is the power law fit to the optical and UV data, including
1-$\sigma$ uncertainties. The red circles are observational data
points, only central values are included. All the observational data
are from Kaplan et al. (2011). The gap between optical/UV and X-ray
is due to neutral hydrogen absorption. The dashed line is RCS model
calculation (Tong et al. 2010).}
\label{1856}
  \end{figure}

\begin{figure}[!htbp]
\centering
\begin{minipage}{0.45\textwidth}
\includegraphics[height=0.3\textheight]{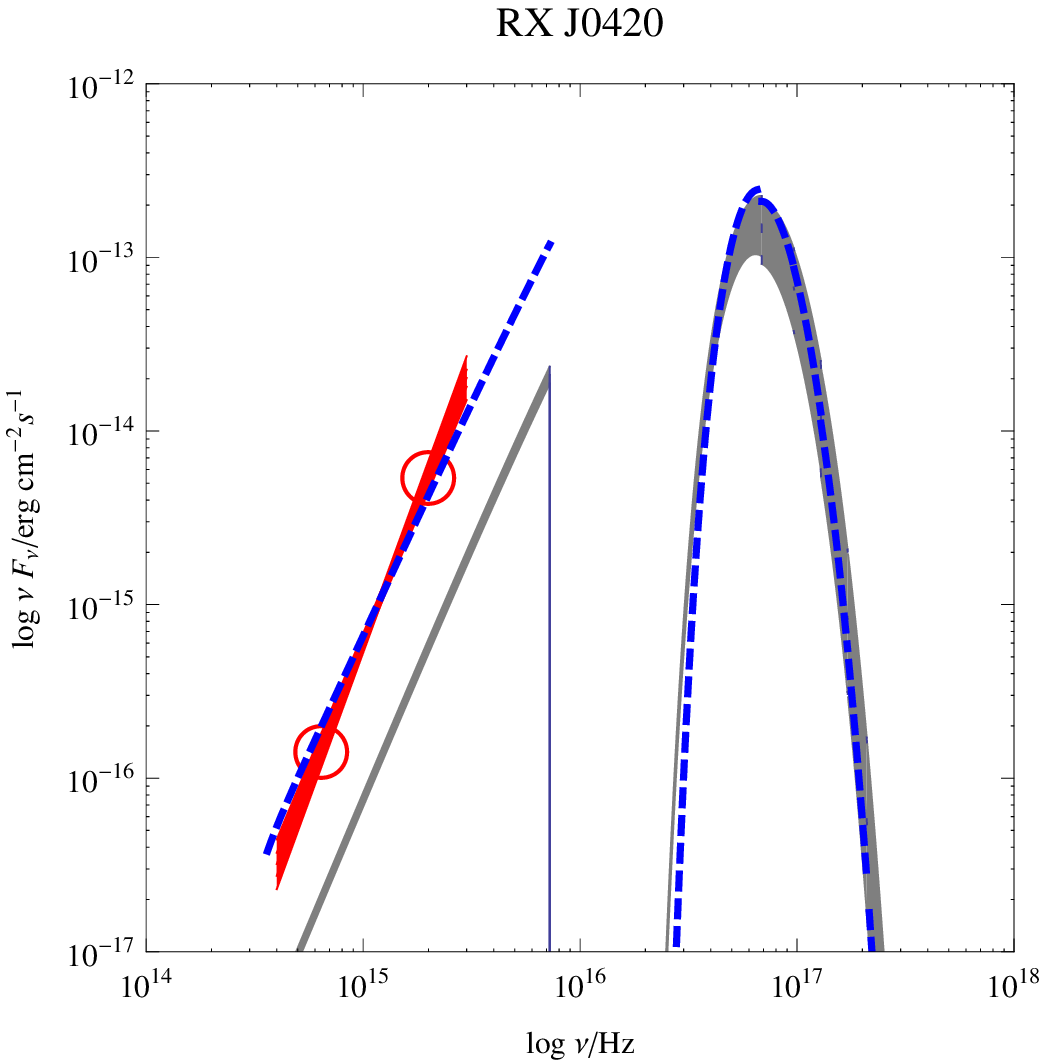}
\end{minipage}
\begin{minipage}{0.45\textwidth}
\includegraphics[height=0.3\textheight]{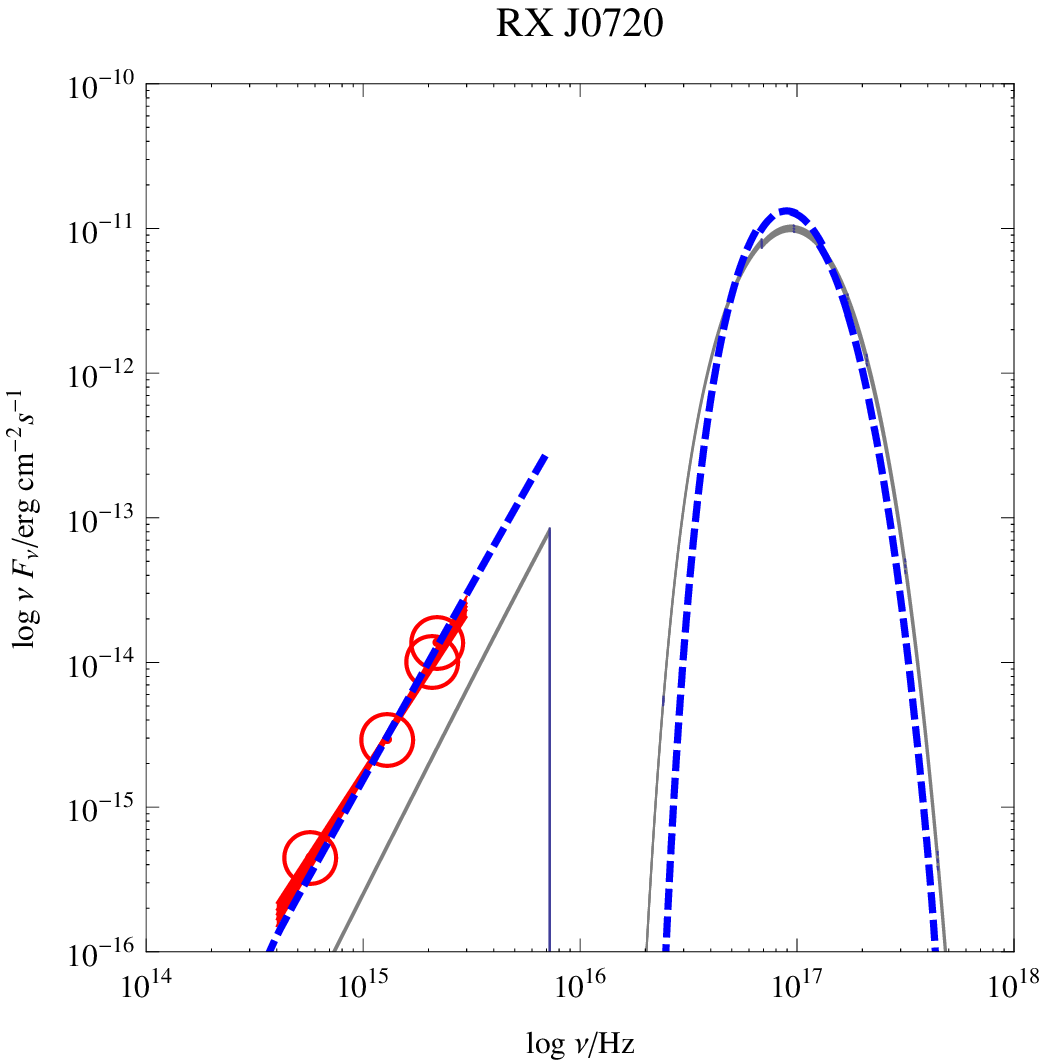}
\end{minipage}

\begin{minipage}{0.45\textwidth}
\includegraphics[height=0.3\textheight]{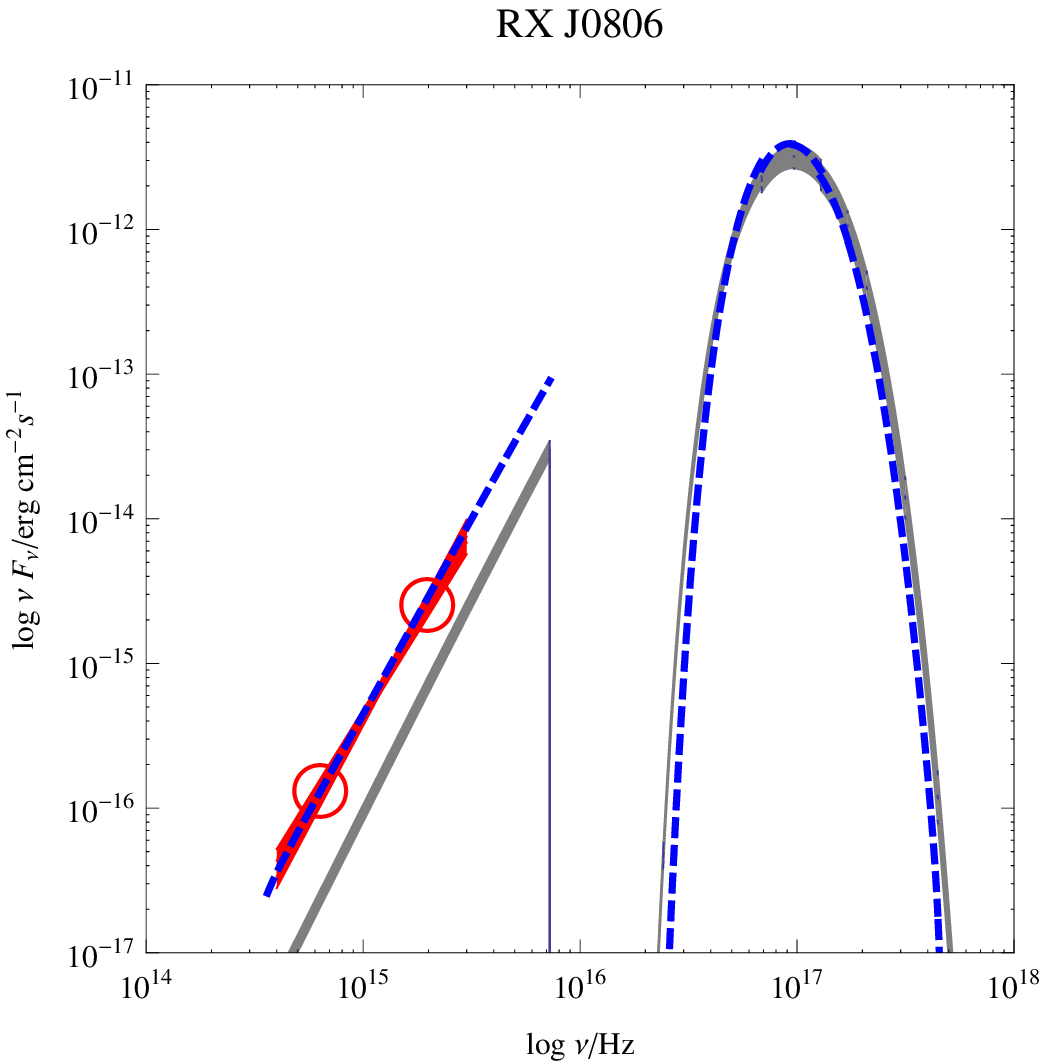}
\end{minipage}
\begin{minipage}{0.45\textwidth}
\includegraphics[height=0.3\textheight]{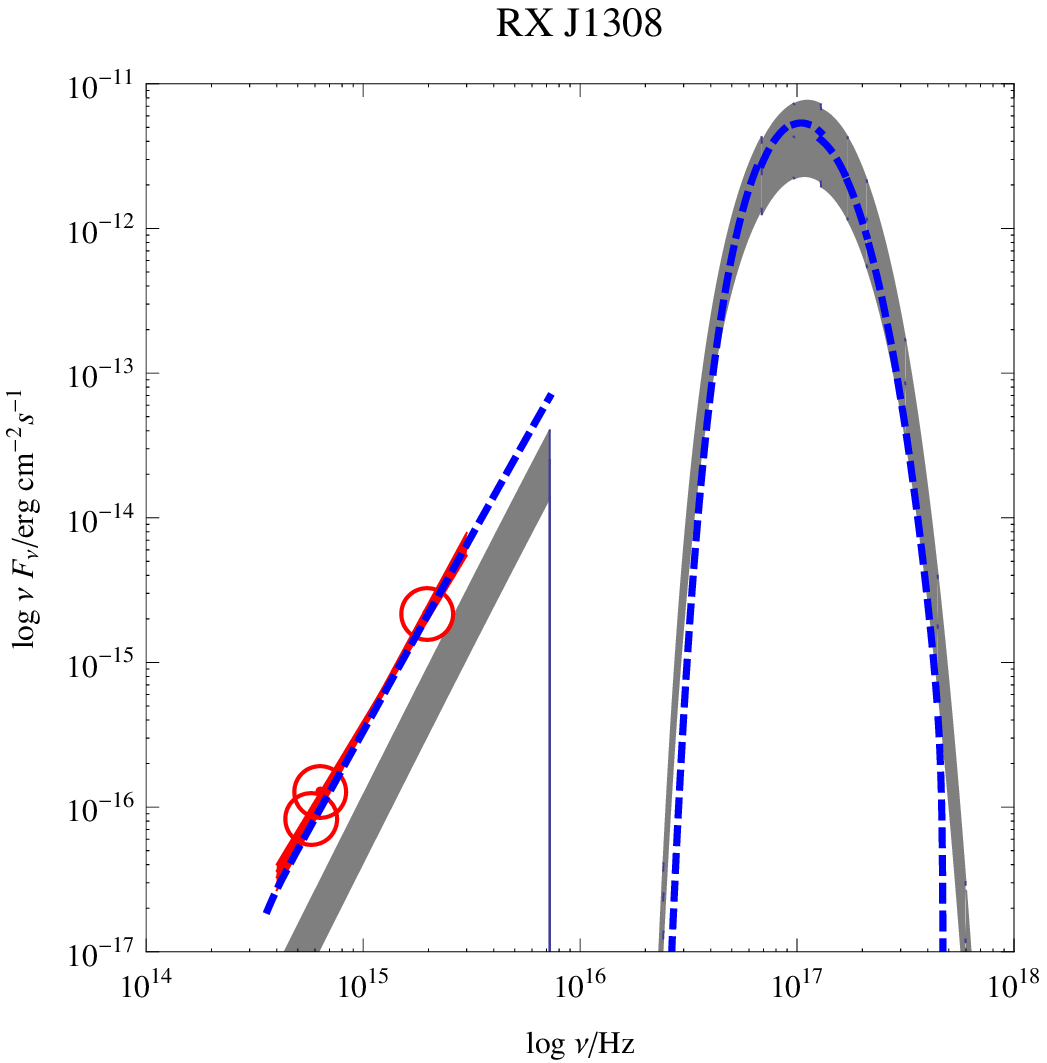}
\end{minipage}

\begin{minipage}{0.45\textwidth}
\includegraphics[height=0.3\textheight]{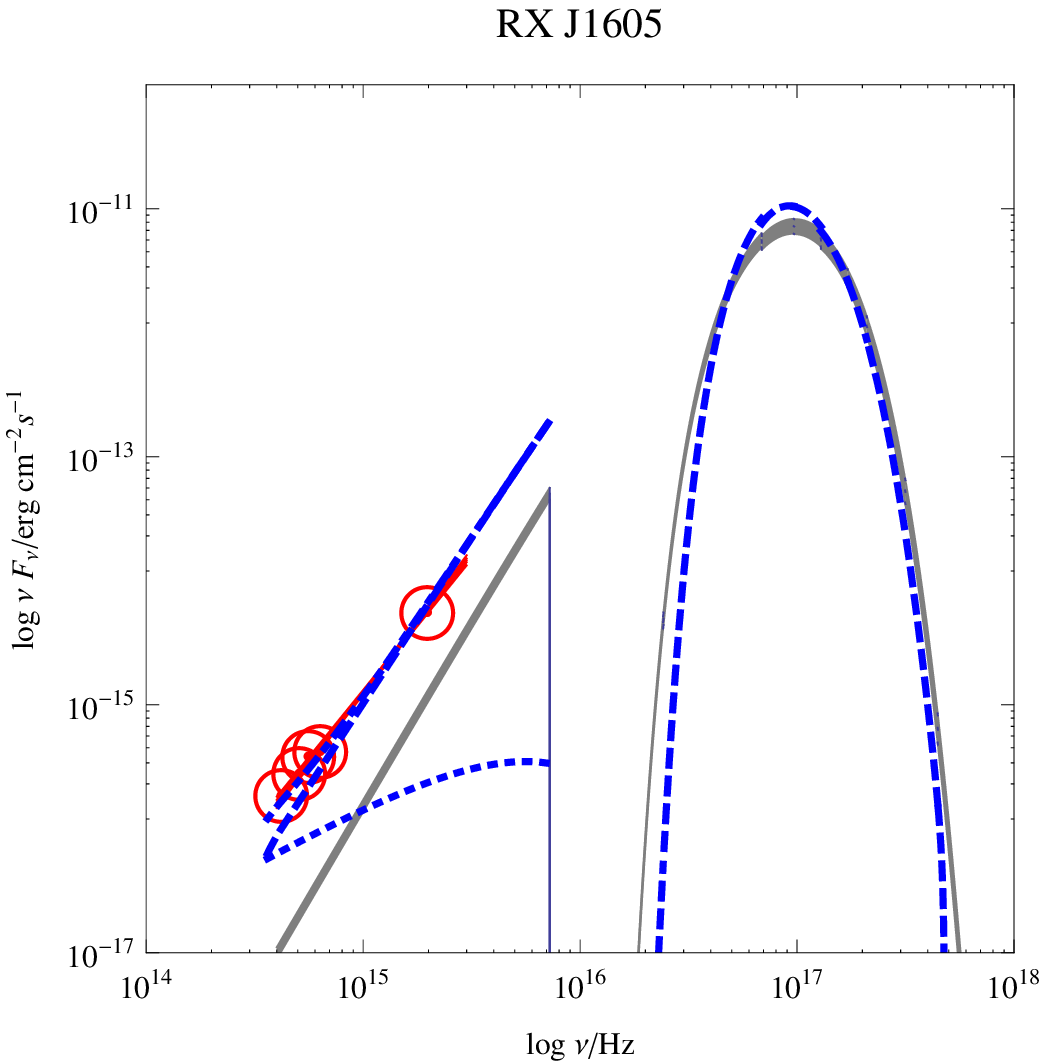}
\end{minipage}
\begin{minipage}{0.45\textwidth}
\includegraphics[height=0.3\textheight]{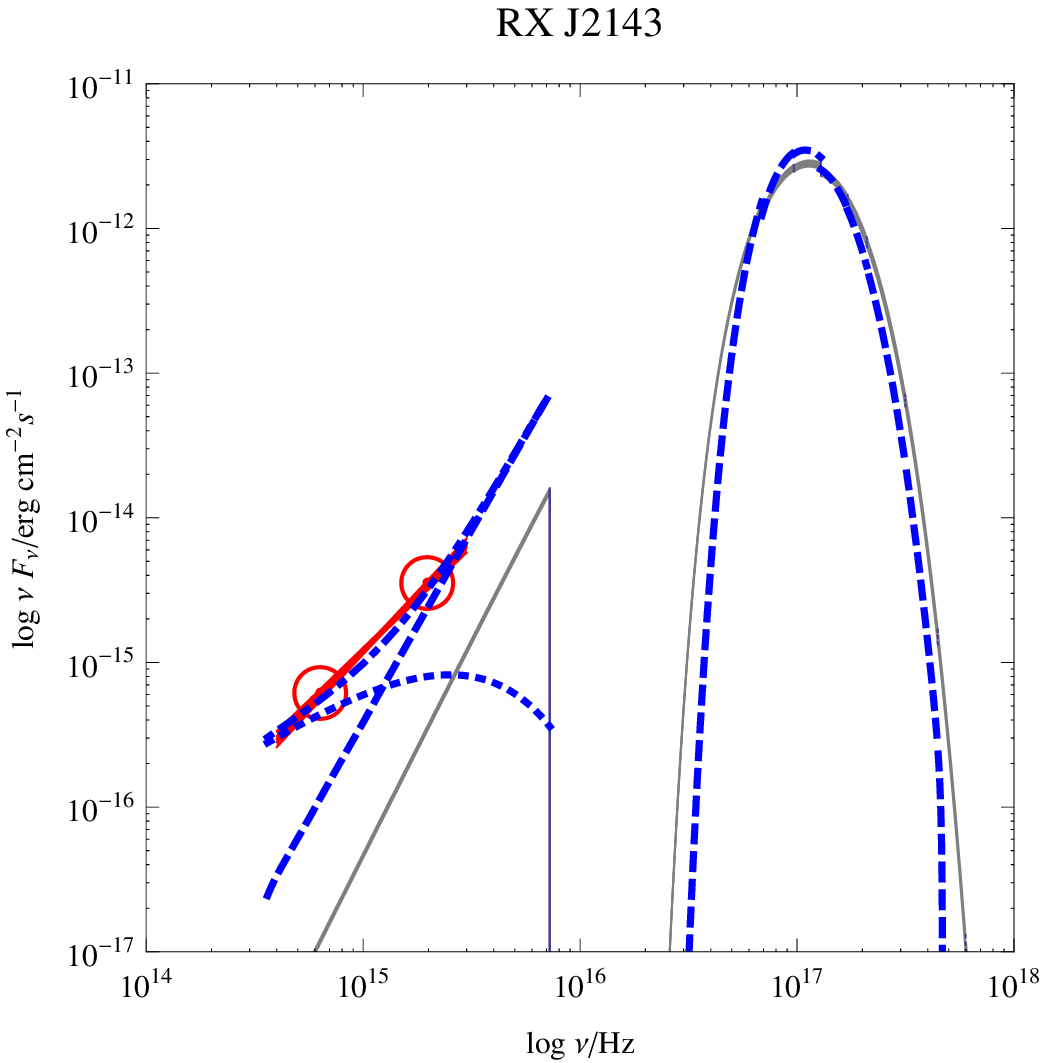}
\end{minipage}

\caption{Calculations for RX J0420.0-5022, RX J0720.4-3125, RX J0806.4-4123,
RX J1308.6+2127, RX J1605.3+3249 and RX J2143.0+0654. The same as figure 2. All observational data are from Kaplan et al. (2011).
The dashed line is the RCS model calculation (Tong et al. 2010). The dotted line is the contribution from electron thermal bremsstrahlung.
The dot-dashed line is the sum of bremsstrahlung and RCS component. The bremsstrahlung component is needed only in the case of
RX J1605.3+3249 and RX J2143.0+0654.}
\label{others}
\end{figure}

The application of RCS model to RX J1856.5-3754 is shown in figure \ref{1856}.
Model calculations for the rest six sources are shown in figure \ref{others}. The input parameters of RCS model
are given is table 1. The neutron star surface temperature $T_{\mathrm{x}}$ is chosen as the temperature of the X-ray blackbody 
(from table 2 in Kaplan et al. 2011)
and is kept fixed during the fitting process. The normalization is the radius of the radiation region over the source distance.
For example, for RX J1856.5-3754, $R/d=10/0.16(\mathrm{km/kpc})$. It means that for a source distance $0.16\,\mathrm{kpc}$
the radius of the radiation region is $10\,\mathrm{km}$. All distance data are from table 4 in Kaplan \& van Kerkwijk (2009).
The electron temperature $T_{\mathrm{e}}$ is about $0.5\,T_{\mathrm{x}}$ or lower. The electron number density $N_{\mathrm{e}}$ is around
$1.5\times 10^{12}\,\mathrm{cm}^{-3}$. The neutral hydrogen column density $N_{\mathrm{H,rcs}}$ is one and a half or two times that of blackbody
fit to the X-ray data. Considering that for all seven XDINSs, their surface magnetic field,
surface temperature, radiation region, X-ray luminosity, and distance are all similar to each other (Kaplan \& van Kerkwijk 2009).
Meanwhile, they all have an optical/UV excess of order ten etc (Kaplan et al. 2011). Hence, these seven sources are similar to each other
observationally. As a consequence of this, the RCS model parameters for these sources are also similar to each other.

The two sources RX J1605.3+3429 and RX J2143.0+0654 have flat optical/UV spectra (especially for RX J2143.0+0654). For most sources,
a power law fit to the optical/UV data gives $F_{\nu} \propto \nu^{\alpha}$, with $\alpha \sim 2$ implying a thermal origin.
While for RX J2143.0+0654, $\alpha \sim 0.5$ (table 5 in Kaplan et al. 2011). The flat spectrum of RX J2143.0+0654 indicates that
nonthermal processes are involved in its optical/UV emission. Noting that the bremsstrahlung emission has a flat spectrum, we propose
that the flat spectrum of RX J2143.0+0654 may due to thermal bremsstrahlung emission of the electron system in addition to the RCS process.

From table 1, the temperature and density of the electron system are
known. Once its radial extension is given, we can calculate its
bremsstrahlung flux at earth (Rybicki \& Lightman 1979). The outer
radius of the electron system $r_{\mathrm{out}}$ determines the
relative importance of the bremsstrahlung component. For RX
J1605.3+3429 and RX J2143.0+0654, $r_{\mathrm{out}}$ is determined
in order to obtain a flat optical spectrum. For the rest five sources, no
bremsstrahlung component is needed. Therefore, an upper bound on
$r_{\mathrm{out}}$ is given. The rotation periods of XDINSs are
about $10\,\mathrm{sec}$. Therefore, the radial extension of the
electron system is at most five percent of the light cylinder
radius, i.e., near the central star. This is why we call it ``pulsar inner radiation belt''. 

When the electron number density varies with radius, e.g,
$N_e(r)=N_e\times(r/R_{\mathrm{ns}})^{-\alpha}$, the fitting results
will be different from the uniform density case as considered above.
In order for electrons in a large space range all contributing to
the bremsstrahlung process, the power index $\alpha$ should not be
significantly larger than 1, i.e., $0<\alpha<1$ is required for RX
J1605.3+3249 and RX J2143.0+0654. While for the rest five sources,
we do not see such a bremsstrahlung component. This is due
to either a small radial extension of the electron system or the electron
number density there decreases with radius very rapidly, i.e.,
$\alpha>1$. The formation of the electron system is discussed in the
appendix.

\begin{table}[!htbp]
  \centering

  \caption{Input Parameters of the RCS Model. Column one to seven are source name, neutron star surface temperature,
normalization, electron temperature, electron density, neutral hydrogen column density, and outer radius of the electron system, respectively.
See text for details.
}
  \label{RCSfit}

  \begin{tabular}{ccccccc}
    \hline\hline
    Source & $T_{\mathrm{x}}$ & R/d & $T_{\mathrm{e}}$ & $N_{\mathrm{e}}$ &  $N_{\mathrm{H,rcs}}$ & $r_{\mathrm{out}}$
     \\
      & ($\mathrm{eV}$) & (km/kpc) & ($\mathrm{eV}$) & ($10^{12}\, \mathrm{cm}^{-3}$) & ($10^{20}\, \mathrm{cm}^{-2}$) & ($10^9\,\mathrm{cm}$)
     \\\hline
    J0420 & 45.0 & 9/0.345 & 22.5 & 2.0 & 3.0 & $<1.2$ \\ \hline

    J0720 & 88.4 & 9/0.36  & 35.4 & 1.6 & 2.1 & $<2.2$ \\\hline

    J0806 & 87.2 & 3.5/0.25 & 43.6 & 1.6 & 2.6 & $<1.2$ \\\hline

    J1308 & 102 & 5.5/0.5 & 51 & 1.4 & 2.7 & $<1.8$ \\\hline

    J1605 & 92.6 & 7.5/0.39 & 23.2 & 1.8 & 2.0 & 1.2 \\\hline

    J1856 & 63.5 & 10/0.16 & 25.4 & 1.8 & 1.7 & $<1.5$ \\\hline

    J2143 & 104 & 4.5/0.43 & 10.4 & 1.5 & 4.6 & 2.2 \\\hline

      \end{tabular}
\end{table}

\section{Discussions and conclusions}

In the discussion section of Kaplan et al. (2011), they also
mentioned the RCS process, which they refer to the paper of Lyutikov
\& Gavriil (2006). However, as pointed in Tong et al. (2010), the
one dimensional treatment of Lyutikov \& Gavriil (2006) can only
result in up scattering. While the optical/UV excess of XDINSs is of
``soft excess'' in nature, which requires down scattering process.
This is the reason that Tong et al. (2010) modeled the RCS process
three dimensionally and employed the Kompaneets equation method.

The RCS model can explain the spectral energy distributions of XDINSs in a wide parameter space. Meanwhile, since the surface X-ray photons
are scattered by a shell of electrons, this will result in a low pulsation amplitude naturally. The X-ray spectra are Planck-like with no
high energy tails. This may imply that XDINSs are possibly quark stars (Xu 2002,2009). At the same time, the possible features may be
proton cyclotron lines or electron cyclotron lines far in the magnetosphere (Turolla 2009).

From figure 3 in Kaplan et al. (2011) we see that
for RX J2143.0+0654 we have only two data points at present. 
Future more optical and UV data will tell us whether its spectrum is really so
flat or not. Moreover, with more data points we can know whether there is curvature in the optical/UV spectrum and whether
there are more spectra components in addition to the power law component. 
The electron thermal bresstrahlung component has a flat spectrum in the IR/optical range, 
with a flux about $0.1\mu \mathrm{Jy}$ for RX J2143.0+0654. 
This flux is much lower than the $H$-band flux upper limit 
($1.6\mu \mathrm{Jy}$ at $1.65 \,\mu m$, Lo Curto et al. 2007). 
Therefore in order to verify whether the flat spectrum of RX J2143.0+0654 
is due to thermal bremsstrahlung process, deeper optical/UV and IR observations are required.

The electron system cools down via the bremsstrahlung
process (possibly also via free-bound transitions, i.e.,
recombination process). Therefore, the central star has to
continually heat the electron system and ionize it. This is similar
to the ``Str$\ddot{\mathrm{o}}$mgren sphere'' around early-type stars (Dyson \&
Williams 1980). The size of the ionized region is $(3S/4\pi N_e^2
\beta_2)^{1/3}$ (Dyson \& Williamas 1980), where $S$ is the ionizing
photon number flux of the central star, $N_e$ is the electron number
density (assuming fully ionized hydrogen plasma), $\beta_2$ is the
recombination coefficient. Similar calculations show that the size
of the ionized region is $\sim 10^{10}\,\mathrm{cm}$, to order of
magnitude consistent with the outer radius used in table 1.

\section*{Acknowledgments}
This work is supported by the National Natural Science Foundation
of China (Grant Nos. 10935001, 10973002), the National Basic Research Program of China
(Grant No. 2009CB824800), and the John Templeton Foundation.

\appendix

\section{Formation of the electron system}

The electron system may be formed similarly to the ``injection through ionization'' process (Luo \& Melrose 2007).
Assuming a neutron star whose mass is $1.4\,M_{\odot}$, radius $10\,\mathrm{km}$ and spatial velocity $100\,\mathrm{km}\,\mathrm{s}^{-1}$, 
it will accrete interstellar medium or circumpulsar material at a rate $\dot{M}=2\pi(G M)^2 V^{-3} \rho_{\mathrm{ISM}} \sim 2\times 10^8 \,\mathrm{g}
\,\mathrm{s}^{-1}$ (Frank et al. 2002), where the density of interstellar medium is assumed to be $10^{-24}\,\mathrm{g}\,\mathrm{cm}^{-3}$. 
Since the accretion rate is rather low, the corresponding Alfv$\acute{\mathrm{e}}$n radius $r_{\mathrm{m}}\sim 4\times 10^{11}\,\mathrm{cm}$ is
larger than the light cylinder radius and corotation radius. 
Only a small fraction of the accreted matter can fall onto the neutron star through diffusion process, similar to the propeller case (Ertan et al. 2007).
Since we are considering diffusion process along magnetic field lines, the diffusion properties are not affected
by the presence of magnetic field. Assuming the plasma deflection length as the typical mean free path (Frank et al. 2002), 
the particle flux of the diffusion process
is $F=D |\triangledown n_{\mathrm{ISM}}| \sim \frac13 \lambda v_{\mathrm{rms}} \frac{n_{\mathrm{ISM}}}{r_{\mathrm{m}}} \sim 10^6 \, 
\mathrm{cm}^{-2} \,\mathrm{s}^{-1}$, 
where $D$ is diffusion coefficient, $\lambda$ the mean free path $\sim 10^{12} \,\mathrm{cm}$, 
$v_{\mathrm{rms}}$ the typical particle velocity $\sim 10 \,\mathrm{km} \,\mathrm{s}^{-1}$,
and $n_{\mathrm{ISM}}$ the interstellar medium number density $\sim 1 \,\mathrm{cm}^{-3}$. 
With such a diffusion flux, the corresponding accretion rate is about two hundred times smaller than the total Bondi-Hoyle accretion rate given above. 
The time scale for the formation of the electron system is about $4\times 10^{3} \,\mathrm{yr} \ll \tau_c$
, where $\tau_c$ is the pulsar characteristic age $\approx 4\times 10^{6} \,\mathrm{yr}$ for RX J2143.0+0654. 
When the diffused plasma comes to the vicinity of the neutron star, they will diffusion across magnetic field lines 
and enter into the pulsar closed field regions. The plasma system of the RCS model may be formed.
The maximum particle number density that can be 
confined by the magnetic field at $r_{\mathrm{out}}$ is (i.e., the kinematic energy density and magnetic one are in equilibrium): 
$B(r_{\mathrm{out}})^2/(8\pi k T_{\mathrm{e}}) \sim 10^{16} \,\mathrm{cm}^{-3} 
\gg N_{\mathrm{e}}$. The plasma system can be confined by the magnetic field.




\label{lastpage}


\begin{thebibliography}{99}
%

\bibitem{Dyson}
Dyson, J. E., Williams, D. A. 1980, Physics of the interstellar medium, Manchester University Press, Manchester

\bibitem{Ertan (2007)}
Ertan, U., Erkut, M. H., Eksi, K. Y., Alpar, M. A. 2007, ApJ, 657, 441

\bibitem{Frank}
Frank, J., King, A., Raine, D. 2002, Accretion power in
astrophysics, Cambridge University Press, Cambridge

\bibitem{Kaplan}
Kaplan, D. L., van Kerkwijk, M. H. 2009, ApJ, 705, 798

\bibitem{Kaplan2011}

Kaplan, D. L., Kamble, A., van Kerkwijk, M. H., Ho, W. C. G. 2011,
ApJ, 736, 117

\bibitem{LoCurto}
Lo Curto, G., Mignani, R. P., Perna, R., Israel, G. L. 2007, A\&A, 473, 539

\bibitem{Luo}
Luo, Q., Melrose, D. 2007, MNRAS, 378, 1481

\bibitem{Lyutikov2006}
Lyutikov, M., Gavriil, F. P. 2006, MNRAS, 368, 690

\bibitem{Ruderman2003}
Ruderman, M. 2003, arXiv: astroph/0310777

\bibitem{Rybicki}
Rybicki, G. B., Lightman, A. P. 1979, Radiative process in
astrophysics, John Wiley \& Sons, New York

\bibitem{Tong2007}
Tong, H., Peng, Q. H. 2007, ChJAA, 7, 809

\bibitem{Tong2008}
Tong, H., Peng, Q. H., Bai, H. 2008, ChJAA, 8, 269

\bibitem{Tong2010}
Tong, H., Xu, R. X., Peng, Q. H., Song, L. M. 2010, RAA, 10, 553

\bibitem{Turolla}
Turolla, R. 2009, in Neutron stars and pulsars, Astrophysics and Space Science Library, 357, 141

\bibitem{van Kerkwijk}
van Kerkwijk, M. H., Kaplan, D. L. 2007, ApSS, 308, 191

\bibitem{Xu2002}
Xu, R. X. 2002, ApJ, 570,  L65

\bibitem{Xu2009}
Xu, R. X. 2009, J.Phys.G: Nucl. Part. Phys., 36, 064010

\end{thebibliography}
\end{document}